\documentclass[12pt,epsf]{article}
\usepackage[xdvi]{graphicx}
\newcommand{\beq}{\begin{equation}}
\newcommand{\eeq}{\end{equation}}
\newcommand{\bea}{\begin{eqnarray}}
\newcommand{\eea}{\end{eqnarray}}

\def\pe2{p_E^2}

\begin{document}
\newcommand{\mpl}{M_{\mathrm{Pl}}}
\setlength{\baselineskip}{18pt}
\begin{titlepage}
\begin{flushright}
OU-HET 598/2008 \\
\end{flushright}

\vspace*{1.2cm}
\begin{center}
{\Large\bf Meta-stable SUSY Breaking
 Model in Supergravity}
\end{center}
\lineskip .75em
\vskip 1.5cm

\begin{center}
{\large Naoyuki Haba }\\

\vspace{1cm}

{\it Department of Physics, Graduate School of Science, Osaka
 University, \\
 Toyonaka, Osaka 560-0043, Japan}\\

%
%
\vspace*{10mm}
{\bf Abstract}\\[5mm]
{\parbox{13cm}{\hspace{5mm}

We analyze a supersymmetry (SUSY) breaking 
 model proposed by 
 Intriligator, Seiberg and Shih 
 in a supergravity (SUGRA) framework. 
This is a simple and natural setup which demands 
 neither extra superpotential interactions nor 
 an additional gauge symmetry.  
In the SUGRA setup, 
 the $U(1)_R$ symmetry is explicitly 
 broken by the constant term 
 in the superpotential, and 
 pseudo-{moduli} field  
 naturally takes non-zero vacuum expectation value 
 through  a vanishing cosmological constant 
 condition. 
Sfermions tend to be heavier than 
 gauginos, and 
 the strong-coupling scale is 
 determined once a ratio of 
 sfermion to gaugino masses 
 is fixed. 

}}

\end{center}
\end{titlepage}
\section{Introduction} 

Supersymmetry (SUSY) is the most promising 
 candidate beyond the standard model (SM). 
The SUSY must be broken in our real world 
 so that an investigation of the SUSY breaking 
 mechanism is important. 
An idea of dynamical SUSY breaking is one of the 
 most attractive scenario\cite{Witten1}, 
 which  must have the  
 global $U(1)_R$ symmetry\cite{Nelson:1993nf}.\footnote{
A non-generic superpotential can also 
 break SUSY dynamically, which will not be considered 
 in this paper.  
} 
However, this $R$-symmetry should be 
 explicitly broken in order to realize 
 the gaugino masses as well as 
 avoid massless $R$-axion. 
And besides, 
 the dynamical SUSY breaking  
 demands 
 complicated chiral gauge 
 theories\cite{Witten2}.\footnote{
If massless matters 
 are
 included, 
 non-chiral 
 theories 
 can  
 also 
 break 
 SUSY 
 dynamically\cite{IT}.  
} 
If we have a possibility to construct 
  the simplest  
 SUSY breaking model,  
 it would be     a non-chiral gauge theory      
 and the 
 SM gauge group is 
 embedded into 
 a subgroup of its flavor symmetry.  
It is a direct    
 gauge mediation of the SUSY breaking, 
  which 
 can suppress 
  the SUSY flavor changing neutral currents 
 (FCNCs)\cite{GM, DNS}. 
However, in order to construct such a model, 
 we must find 
 a dynamical SUSY breaking model in non-chiral gauge
 theory without massless $R$-axion.\footnote{
There is another difficulty of an existence of 
 too many fields with the SM quantum numbers.  
} 
This task seems almost impossible due to 
 the theorems in Refs.\cite{Witten1, Nelson:1993nf}.

This situation is drastically changed 
 if we give up an ordinary sense that we are living 
 in the true vacuum. 
Recently, Intriligator, Seiberg and Shih (ISS) have discovered 
 a meta-stable SUSY breaking vacuum
 in ${\cal N}=1$ non-chiral SUSY gauge theory 
 in a free magnetic phase\cite{ISS}.  
The model has $SU(N_c)$ gauge group 
 with massive $N_f$ fundamental and anti-fundamental
 chiral-superfields   
  in the range of 
 $N_c < N_f < \frac{3}{2}N_c$\cite{Seiberg}. 
Since the SUSY breaking vacuum is not a global 
 minimum but a meta-stable vacuum, 
 the existence of the $R$-symmetry is not
 necessarily required as well as the 
 theory can be non-chiral. 
This situation is attractive as long as 
 the meta-stable vacuum is long-lived 
 compared with an age of the universe, and    
 then a lot of  
 researches on the meta-stable SUSY  
 breaking have been done 
 in various aspects\cite{sample}-\cite{AKO}. 
However, from the view point of 
 phenomenology, 
 there exists still a difficulty 
 for generating 
 the suitable 
 gaugino masses in the ISS model.  
For this purpose, people 
 have introduced 
 explicit $R$-symmetry breaking interactions 
 in the superpotential by hand\cite{KOO, KOO2} 
 and also 
  an additional gauge symmetry\cite{CST}.  
These extensions of the ISS model 
 seem complicated and artificial.

In this paper 
 we analyze the ISS 
 model 
 in a supergravity (SUGRA) framework.\footnote{
$R$-axion obtains a mass in the SUGRA 
 framework\cite{brp}. 
And, the ISS model in the SUGRA setup 
 was also considered in Ref.\cite{AKO1, AKO}. 
} 
This is a simple and natural extension which demands 
 neither extra superpotential interactions nor 
 an additional gauge symmetry.  
In the SUGRA setup, 
 the $R$-symmetry is explicitly 
 broken by the constant term 
 in the superpotential, and 
 pseudo-{moduli} field 
 naturally takes non-zero vacuum expectation value (VEV) 
 through  a vanishing cosmological constant 
 condition. 
Sfermions tend to be heavier than 
 gauginos, and 
 the cutoff scale of the magnetic 
 description is 
 determined once a ratio of 
 sfermion to gaugino masses 
 is fixed. 
We will also show 
 the meta-stable SUSY 
 breaking vacuum can be sufficiently
 long-lived. 

\section{ISS model}

First let us show 
 a basic idea of the ISS model.  
The model is described as
 an ${\cal N}=1$ SUSY $SU(N_f-N_c)$ gauge theory 
 which consists of 
 $N_f$ dual quarks $q$, $\tilde{q}$ 
 and a gauge singlet $M$.
This is a magnetic dual of  
 SUSY $SU(N_c)$ gauge theory with massive $N_f$ flavors. 
Superpotential and K\"ahler potential 
 are 
 given by
\bea
{W} &=& q M \tilde{q} - {\rm Tr} [m^2 M], 
\label{1} \\
K_0 &=& {\rm Tr} [M^\dag M] + q^\dag q + \tilde{q}^\dag \tilde{q}, 
\label{2}
\eea
respectively. 
Trace is taken in the flavor space. 
This K\"ahler potential is a canonical form 
 since this model is an IR free theory. 
There exists 
 the $R$-symmetry, where 
 $R$-charge assignment is 
 $R(M)=2$ and $R(q)=R(\tilde{q})=0$. 
We decompose  $M$, $q$, $\tilde{q}$,
 and take $m$ as                              
\bea
&&M = 
\left(
 \begin{array}{cc}
\hat{Y}_{ab} & Z_{aB} \\
\tilde{Z}_{Ab} & \hat{\Phi}_{AB} \\
\end{array}
\right), \quad 
q = 
\left(
\begin{array}{c}
\chi_a \\
\rho_A \\
\end{array}
\right), \quad 
\tilde{q} = 
\left(
\begin{array}{c}
\tilde{\chi}_a \\
\tilde{\rho}_A \\
\end{array}
\right), \quad 
m = 
\left(
\begin{array}{cc}
m \delta_{ab} & 0 \\
0 & \tilde{m} \delta_{AB} \\
\end{array}
\right) 
\eea
where $a,b = 1, \cdots, N_f-N_c$ and $A,B=1,\cdots,N_c$ 
 are flavor index.   
Then Eqs.(\ref{1}) and (\ref{2}) 
 are rewritten by the components as 
\bea
{W} &=& \chi \hat{Y} \tilde{\chi} +\chi Z \tilde{\rho} 
+ \tilde{\chi} \tilde{Z} \rho + \rho \hat{\Phi} \tilde{\rho} 
- m^2 {\rm Tr}[\hat{Y}] - \tilde{m}^2 {\rm Tr}[\hat{\Phi}] ,
\label{spotdeco}\\
K_0 &=& {\rm Tr} \left[|\hat{Y}|^2
 + |Z|^2 + |\tilde{Z}|^2 + |\hat{\Phi}|^2\right] 
+ |\chi|^2 + |\rho|^2 + |\tilde{\chi}|^2 + |\tilde{\rho}|^2. 
\label{kahlerdeco}
\eea
By using the field redefinitions, 
 we can always take $\rho = \tilde{\rho} = 0$.  
And, $F$-flat conditions, 
 ${\partial W}/{\partial \chi}=
 {\partial W}/{\partial \tilde{\chi}}=0$,  
 are satisfied 
 in the direction of 
 $\hat{Y}=Z=\tilde{Z} = 0$. 
Then 
 the remaining 
  $F$-flatness conditions are given by 
\bea
&& \frac{\partial W}{\partial \hat{Y}} 
= \chi \tilde{\chi} - m^2 \delta_{ab}, \;\;\;
 \frac{\partial W}{\partial \hat{\Phi}}
  = -\tilde{m}^2 \delta_{AB}, 
\label{7}
\eea
which show  
 that the minimum exists at  
 $\chi \tilde{\chi} = m^2 \delta_{ab}$. 
Since the trace part of $\hat{\Phi}$, which we denote 
 $\Phi \equiv {1\over N_c}{\rm Tr}[\hat{\Phi}]$, 
 has non-zero $F$-term $\tilde{m}^2$, 
  the SUSY   is spontaneously broken. 
The $\Phi$ is a pseudo-moduli field whose 
 fermionic component is the Nambu-Goldstone (NG) 
 fermion of the spontaneous SUSY breaking.  
We should notice that the traceless part of 
 $\hat{\Phi}$, which we denote 
 $\Phi_0 \equiv \hat{\Phi}-\Phi$, 
 has no $F$-term due to the absence of 
 tadopole term. 
In this vacuum,  
 the gauge symmetry $SU(N_f-N_c)$ is 
 completely broken, and   
 the flavor symmetry is reduced to 
 $SU(N_f-N_c)\times SU(N_c) \times U(1)_B$.

$\hat{\Phi}$ and $\chi-\tilde{\chi}$ 
 remain massless at the tree level while other
 fields obtain masses of ${\cal O}(m)$.\footnote{
$\chi-\tilde{\chi}$ is a NG 
 superfield from 
 the broken $U(1)_B$, which could be absorbed into 
 the vector super-multiplet 
 by gauging the $U(1)_B$. 
 } 
This means that the VEV of $\hat{\Phi}$ 
 cannot be determined in the tree level. 
Since 
 the gaugino masses are never generated unless 
 both the SUSY and R-symmetry are broken, 
 we need $\Phi \neq 0$. 
However, even if $\Phi \neq 0$
  and non-zero gaugino masses are generated 
 through the 
 quantum corrections,  
 the $R$-symmetry is spontaneously broken 
 which induces an unwanted massless $R$-axion. 
Therefore,                                            
 we need an 
 explicit $R$-symmetry breaking. 
What is the most natural setup? 
The answer 
 might  be the SUGRA,  
 in which $\Phi$ 
 obtains 
 a non-zero VEV as will be shown later.\footnote{
Real part of the scalar component
 of $\chi-\tilde{\chi}$ also obtains a 
 mass (from corrections of higher order K\"{a}hler potential) 
 even when $U(1)_B$ is
 not gauged, where  
 its imaginary part is a NG boson  and  
 a fermion component  
 is still massless. 
} 
In fact the ISS model has already implied 
 the existence of SUGRA, since 
 the massless NG fermion 
 should be 
 absorbed into the longitudinal mode of the gravitino. 
The $R$-symmetry is explicitly broken 
 in the SUGRA framework  
 through
 the constant term of the superpotential, 
 which plays a crucial role for realizing 
 the vanishing
 cosmological constant.
We will show that this setup 
 demands 
 neither extra superpotential interactions nor 
 an additional gauge symmetry 
 differently from models so far\cite{KOO, KOO2, CST}. 

\section{ISS model in SUGRA}

Let us now 
  consider the ISS model 
 in the SUGRA setup. 
We also introduce the next leading order of 
 the K\"{a}hler potential\footnote{
The coefficients of ${(q^\dag q)^2/\Lambda^2}$ and 
 ${(\tilde{q}^\dag \tilde{q} )^2/\Lambda^2}$ can be 
 (of cause) different. 
In this case the following analyses a little bit change, 
 but which are easily calculated. 
Here we just take the same coefficient 
 for simplicity. 
}  
\bea
K_1 &=& - \lambda
 {{\rm Tr} [(M^\dag M)^2] \over \Lambda^2}
 - \lambda_a 
 {({\rm Tr} [M^\dag M])^2 \over \Lambda^2}
  - \lambda' \left({(q^\dag q)^2\over \Lambda^2}+
 {(\tilde{q}^\dag \tilde{q} )^2\over \Lambda^2}\right), 
\label{kahler1}
\eea
  for the pseudo-{moduli} 
 not to take a 
 larger VEV than $\Lambda$.\footnote{
If we do not introduce  $K_1$,            
 the VEV of pseudo-{moduli}  would be the Planck scale 
 as the usual Polonyi model\cite{Polonyi}. 
The VEV larger than $\Lambda$ 
 is meaningless in the dual description. 
The similar analyses (including $K_1$ in the Polonyi model)
 had been done
 in Refs.\cite{Kitano}. 
}
Here 
$\Lambda$ is the strong-coupling scale of 
 this theory. 
We assume  
 the negative 
 signs, $\lambda, \lambda' \sim 1$ and  
 $\lambda_a=0$ \footnote{
It is just for simplicity. 
The case of $\lambda_a\neq 0$ is easily calculated,
 where  
 interactions  Tr$[|\hat{Y}|^2 |\Phi|^2]$, 
 Tr$[|\tilde{Z}|^2 |Z|^2]$ 
 are 
   added and 
 the following discussions are changed a little. 
} 
 in Eq.(\ref{kahler1}). 
Then 
 $K_1$ is written 
 in components as 
\bea
K_1 &=& - {\lambda\over\Lambda^2} {\rm Tr} \left[
|\hat{Y}|^4 + |Z|^4 + |\tilde{Z}|^4 + |\hat{\Phi}|^4 
+2 \left(
\hat{\Phi}^\dag \hat{Y}^\dag Z \tilde{Z} +
\hat{\Phi} \hat{Y} Z^\dag \tilde{Z}^\dag \right.\right. \nonumber \\
&& 
\left.\left.
+(|\hat{Y}|^2+|\hat{\Phi}|^2) (|\tilde{Z}|^2+|Z|^2)\right)\right]
-{\lambda' \over \Lambda^2}  
\left[|\chi|^4 + |\rho|^4 + |\tilde{\chi}|^4 +
 |\tilde{\rho}|^4 \right. \nonumber \\
&&\left. +2\left(|\rho|^2|\chi|^2+
|\tilde{\rho}|^2|\tilde{\chi}|^2\right)\right].
\label{kahler2}
\eea 
Notice that 
 this contains the term,
 Tr[$\hat{\Phi}^\dag \hat{Y}^\dag Z \tilde{Z}+$h.c.],
 which is absent in the original ISS model.

The scalar potential in the SUGRA
 is given by 
\beq
V=e^{K/M_P^2}
\left\{ F_i^\dag K_{\bar{i}j}^{-1}F_j-{3|W|^2\over M_P^2}\right\},
\eeq
where $M_P=1/\sqrt{8\pi G} \simeq 2.4\times 10^{18}$ GeV
 is
 the reduced Planck scale, 
 and indices mean derivative. 
The 
 $F$-term in the SUGRA is given by 
\beq
F_i^\dag =-{W_i}-K_{i}{{W}\over M_P^2}. 
\eeq

Let us search the potential minimum 
 in the direction of 
 trace part of $M$. 
Although the traceless part of $M$ 
 might 
 take VEVs at their minimum, which 
 can be taken away by the shift 
 of the origin.   
This effect is renormalized by redefining 
   couplings     
 in the following potential.\footnote{  
This is correct up to 
 ${\cal O}(\Lambda^{-2})$ and
 ${\cal O}(M_P^{-2})$. 
The author would like to thank K. Yoshioka 
  for pointing out it. 
}   
The scalar potential 
 in this direction   
 is calculated 
 as
\bea
V&\simeq&e^{K/M_P^2}\; 
\left\{ {\rm Tr} \left[
{1\over1-{4\lambda\over\Lambda^2}|\hat{\Phi}|^2}
\left| \tilde{m}^2 \delta_{AB}
+\hat{\Phi}^\dag\left(1-{2\lambda\over\Lambda^2}|\hat{\Phi}|^2
\right){W\over M_P^2}
\right|^2 \right.\right. \nonumber \\
&&\;\;\;\;\;\;\;\;\;\;\;\;\;\;\;\;
\left.\left.
+{1\over1-{4\lambda\over\Lambda^2}|\hat{Y}|^2}
\left| \chi\tilde{\chi}- m^2 \delta_{ab}+
\hat{Y}^\dag\left(1-{2\lambda\over\Lambda^2}|\hat{Y}|^2
\right){W\over M_P^2}
\right|^2  \right] \right. \nonumber \\
&&\;\;\;\;\;\;\;\;\;\;\;\;\;\;\;\;
\left.\left.
+{1\over1-{4\lambda'\over\Lambda^2}|\chi|^2}
\left| \hat{Y}\tilde{\chi}+
\chi^\dag\left(1-{2\lambda'\over\Lambda^2}|\chi|^2
\right){W\over M_P^2}
\right|^2  \right.\right. \nonumber \\
&&\;\;\;\;\;\;\;\;\;\;\;\;\;\;\;\;
\left.
+{1\over1-{4\lambda'\over\Lambda^2}|\tilde{\chi}|^2}
\left| \hat{Y}{\chi}+
\tilde{\chi}^\dag\left(1-{2\lambda'\over\Lambda^2}|\tilde{\chi}|^2
\right){W\over M_P^2}
\right|^2
 -{3|W|^2\over M_P^2}\right\} .
\eea
The zero-th order 
  ($(1/\Lambda)^0$ and $(1/M_P)^0$) of $V$   
 corresponds to 
 the tree level potential of the ISS model,  
\beq
V_0 = {\rm Tr}\left[ |\tilde{m}^2 \delta_{AB}|^2 +
 |\chi\tilde{\chi}-m^2 \delta_{ab}|^2 \right] +
 |\hat{Y}\chi|^2+|\hat{Y}\tilde{\chi}|^2.   
\eeq 
This determines the vacuum at
\beq
\chi\tilde{\chi}=m^2 \delta_{ab}, \;\;\;\;\; \hat{Y}=0.
\eeq 
The $D$-flat condition shows 
 $|\chi|=|\tilde{\chi}|$.
Notice again that pseudo-moduli 
$\Phi$ is not determined at the tree level.

The next order of non-zero $V$ is 
  $(1/\Lambda)^2$ and $(1/M_P)^2$. 
By taking $D$-flat conditions with 
 assuming real VEVs of the fields,                                             
 stationary conditions of 
  $\hat{\Phi}$ and $\hat{Y}$ show 
\bea        
\label{19}
&&0={dV \over d\hat{\Phi}}\simeq
{2 \lambda \tilde{m}^4 \hat{\Phi} \over \Lambda^2}-
{m^2\tilde{m}^2 {\rm Tr}[\hat{Y}]
 - c\tilde{m}^2 \over M_P^2}, \\
&&0={dV \over d\hat{Y}}\simeq
2m^2 \hat{Y}+{8 \lambda' m^4 \hat{Y} \over \Lambda^2}-
{2m^2\tilde{m}^2 {\rm Tr}[\hat{\Phi}]
 -(4m^4{\rm Tr}[\hat{Y}]+\tilde{m}^4\hat{Y})-2cm^2 \over M_P^2},
\label{200}
\eea
up to ${\cal O}(\Lambda^{-2})$ and ${\cal O}(M_P^{-2})$. 
Here 
 we take  the condition of 
 $\hat{\Phi}, \hat{Y} \ll \Lambda$ (neglecting
  higher order terms of 
 $\hat{\Phi}^n, \hat{Y}^n, \hat{\Phi}^k \hat{Y}^l$
 ($n\geq 2,\; k,l\geq 1$)),  
 and take 
 $m, \tilde{m}$ as real numbers, for simplicity. 
$c$ is the constant term
 in the superpotential which is meaningless 
 in the global SUSY theory. 
This constant $c$ 
 breaks $R$-symmetry explicitly,
 which 
 plays a crucial role to
 realize vanishing cosmological constant. 
Equations (\ref{19}) and (\ref{200}) suggest 
 that      the (local) minimum exists at 
\beq
\Phi \simeq -{c\Lambda^2\over 2\lambda \tilde{m}^2 M_P^2}, \;\;\;\;
Y \simeq -{c \over M_P^2}, 
\eeq
where $Y \equiv {1\over N_f-N_c}{\rm Tr}[\hat{Y}]$.  
Thus, $\Phi$ is determined and $Y$ is shifted
 by  the SUGRA (and its 
$R$-symmetry breaking) effects. 
Energy scales of the VEVs 
 are surely below the dynamical 
 scale of $\Lambda$. 
The height of the potential at this 
 minimum is given by 
\beq
V_{\rm (min)} \simeq N_c \tilde{m}^4-{3 c^2 
  \over M_P^2}\;, 
\eeq  
where we neglect ${\cal O}(m^2\tilde{m}^4/M_P^2)$ 
 and ${\cal O}(\Lambda^{-2} M_P^{-2})$. 
Thus, 
 $c$ must be 
\beq
c \simeq \sqrt{N_c\over3}\;{\tilde{m}^2 M_P}
\eeq 
in order to realize 
 the  vanishing cosmological constant, 
 $V_{\rm (min)}\simeq 0$.

In summary, the (local) minimum
 exists at 
\bea
\label{20}
&& 
\Phi \simeq -{\sqrt{N_c}\Lambda^2\over 2\sqrt{3}\lambda M_P},\;\;\;\;\;\; 
Y \simeq -\sqrt{N_c\over3}{\tilde{m}^2 \over M_P}, \;\;\;\;\;\; 
\chi=\tilde{\chi} = m, \\
&& 
F_\Phi \simeq \tilde{m}^2, \;\;\;\;\;\; 
F_Y = 0, 
 \;\;\;\;\;\; 
F_\chi =F_{\tilde{\chi}}=0, 
\label{21}
\eea
 up to ${\cal O}(\Lambda^{-2})$ and ${\cal O}(M_P^{-1})$.\footnote{
$\chi$, $\tilde{\chi}$ and $F_Y$ have corrections of 
 ${\cal O}(M_P^{-2})$, while $F_\chi$ and  $F_{\tilde{\chi}}$ 
 have only corrections of ${\cal O}(M_P^{-4})$. 
} 
The 
 eigenvalues of mass matrix
 (curvatures at this minimum) 
 of scalar fields, $\Phi$, $Y$, 
 $\chi$, and $\tilde{\chi}$ 
 are all positive of order 
\beq
{\tilde{m}^4 \over \Lambda^2},\;\; m^2, \;\; m^2, \;\; m^2,
\label{26}
\eeq
up to ${\cal O}(\Lambda^{-2})$,
 respectively. 
Off diagonal elements of the mass matrix 
 are at most ${\cal O}(M_P^{-2})$,
 which can be neglected. 
This means that 
 the (local) minimum is surely (meta-)stable.

\section{SUSY breaking mediation}

In the previous section, 
 a non-zero VEV of pseudo-{moduli} $\Phi$  
 is naturally obtained  
 in the SUGRA framework.
This implies that the gaugino masses      
 can be generated 
 if the SM 
 gauge group is embedded into the unbroken flavor symmetry,  
 $SU(N_c)$ or $SU(N_f-N_c)$. 
$\rho$ and $\tilde{\rho}$ are identified as messengers,
 and their mass matrix     
 is given by 
\bea
W \supset (\rho, Z)
\left(
\begin{array}{cc}
\Phi & m \\
m & \left[{\Phi^\dag Y^\dag\over\Lambda^2}\right]_{\bar{F}} \\
\end{array}
\right)
\left(
\begin{array}{c}
\tilde{\rho} \\
\tilde{Z} \\
\end{array}
\right) 
\equiv 
(\rho, Z)
{\cal M}
\left(
\begin{array}{c}
\tilde{\rho} \\
\tilde{Z} \\
\end{array}
\right). 
\label{23}
\eea
The gaugino masses 
 are given by 
\beq
M_{\lambda_i} = \frac{\alpha_i}{4 \pi}N 
 F_\Phi \frac{\partial}{\partial \Phi}
 \;{\rm log}\;{\rm det}{\cal M} 
\label{g24}
\eeq
where 
 $\alpha_i \equiv g_i^2/(4\pi)$ 
 $(i=SU(3)_c, \;SU(2)_L, \; U(1)_Y)$, and 
 $N$ is a flavor number of messengers as 
 $N_f-N_c$  ($N_c$) when 
 the SM gauge group is embedded into 
 $SU(N_c)$ ($SU(N_f-N_c)$). 

We should notice that Eq.(\ref{g24}) can take non-zero values 
 thanks to the  K\"{a}hler potential $K_1$ 
 in Eq.(\ref{kahler2}), since  
 it 
 contains 
 the interaction, $\Phi^\dag Y^\dag Z \tilde{Z}$. 
Recall that 
 the original ISS model does not have 
 (an $R$-symmetry breaking)
 $Z\tilde{Z}$ mass term in the superpotential, then 
 it is difficult to produce gaugino 
 masses. 
So this situation 
 is expected to be modified in the SUGRA framework. 
However, unfortunately,  
 the gaugino masses    
 are still too tiny as 
\beq
M_{\lambda_i} \sim
 N\sqrt{N_c}\;{\alpha_i \over 4\pi}{\tilde{m}^2\over m^2}
{\tilde{m}^4\over \Lambda^2M_P}
\sim
 N\sqrt{N_c}\;{\alpha_i \over 4\pi}{\tilde{m}^2\over m^2}
{\tilde{m}^2\over\Lambda^2}m_{3/2}, 
\label{25gaugino}
\eeq
{}from Eqs.(\ref{20}) and (\ref{21}), which have shown 
 $[\Phi^\dag Y^\dag/\Lambda^2]_{\bar{F}}
 \simeq \sqrt{N_c}\tilde{m}^4/(\Lambda^2M_P)$.  
Equation (\ref{25gaugino}) 
 suggests that the gaugino masses induced from 
 gauge mediation are smaller than 
 anomaly mediation effects\cite{Randall:1998uk}
 due to $m,\tilde{m}\ll \Lambda$,                         
 where 
 the gravity mediation  dominates the gauge mediation.

Although 
 the suitable gaugino masses 
 are not induced from Eq.(\ref{g24}), 
 we should remember that 
 there are still cubic order contributions
 (${\cal O}(F^3_\Phi)$)
 of SUSY breaking\cite{INTY},  
 which induce the gaugino masses as 
\bea
&&M_{\lambda_i} \simeq N \frac{\alpha_i}{4\pi} 
\left(\frac{F_\Phi}{\Phi^2} \right)^2 
\frac{F_\Phi}{\Phi} 
\sim {N\over N_c^2\sqrt{N_c}} \frac{\alpha_i}{4\pi} 
\frac{\tilde{m}^6 M_P^5}{\Lambda^{10}}.  
\label{gaugino} 
\eea
On the other hand, the sfermion masses are generated 
 by the usual two-loop diagrams as                      
\bea
&&m^2_{\tilde{f}} \simeq N C_i 
\left( \frac{\alpha_i}{4\pi} \right)^2 
\left(\frac{F_\Phi}{\Phi} \right)^2 
\sim 
 {N\over N_c} C_i \left( \frac{\alpha_i}{4\pi} \right)^2 
\frac{\tilde{m}^4 M_P^2}{\Lambda^4}, 
\label{sfermion}
\eea
where
 $C_i$s are the quadratic Casimir coefficients,
 that is, $C_3=4/3$, 
 $C_2=3/4$, and 
 $C_1=(3/5)Y^2$. 
We should notice that Eqs.(\ref{gaugino}) and (\ref{sfermion}) 
 are correct only when $\Phi^2 > F_\Phi$ 
 ($\Lambda^4/M_P^2 > \tilde{m}^2$ from Eq.(\ref{20})), 
 so that  
 we cannot take a (global SUSY)
 limit of $M_P\rightarrow \infty$  
 where $\Phi \rightarrow 0$. 
Equations (\ref{gaugino}) and (\ref{sfermion}) suggest that 
 the sfermion masses are heavier than the gaugino masses 
 as 
\bea
\frac{m_{\tilde{f}}}{M_\lambda} \sim N_c^2 \sqrt{\frac{C_i}{N }}
\left(\frac{\Lambda^2}{\tilde{m}M_P} \right)^4, 
\eea 
thus, this model tends to induce the 
 split-SUSY spectrum\cite{ArkaniHamed:2004fb}. 

Taking the color stability condition $|\tilde{m}|<|\Phi| $ 
 into account, 
 a moderate example of the split-SUSY spectrum 
 might be 
 $m_{\tilde{f}}/M_\lambda\sim 100$,  
 although 
 some tunings of parameters are required 
 for the hierarchy problem. 
Let us analyze a case of 
 $M_\lambda = {\cal O}(100)$ GeV and 
 $m_{\tilde{f}} = {\cal O}(10)$ TeV from 
 now on.\footnote{
A case of 
 $M_\lambda = {\cal O}(100)$ GeV and 
 $m_{\tilde{f}} = {\cal O}(1)$ TeV 
 can be also analyzed in the same way,
 where $N_c, N_f$ factors and ${\cal O}(1)$ coefficients   
 must be  carefully taken into account. 
} 
A rough estimation 
 which neglects $N_c, N_f$ factors and ${\cal O}(1)$ 
 coefficients  
 suggests\footnote{
Hereafter we show absolute values of $m,\tilde{m}$ and VEVs.  
}      
\bea
\tilde{m} \sim 10^{6.5}~{\rm GeV}, \;\;
\Lambda \sim 10^{12.5}~{\rm GeV} \;\;  
(\Phi  \sim 10^{7}~{\rm GeV}),
\label{29}
\eea
where the magnitude   
 of $\Phi$ is really smaller than 
 $\Lambda$ (and also $M_P$).
Notice that all energy scales are
 determined once the ratio of 
 sfermion masses to gaugino masses 
 is fixed.  
In the  case of Eq.(\ref{29}), 
    we can show that the gravity mediation effects
 are much smaller than the gauge mediation ones 
 because 
\bea
m_{3/2} \simeq \frac{F_\Phi}{M_P} 
={\cal O}(10)~{\rm keV} .
\eea
Here $m_{3/2}$ is the gravitino mass,  
 which means the gravitino 
 is the lightest superparticle 
 in this model.\footnote{
${\cal O}(10)$ keV gravitino is free from 
 the 
 gravitino problem, but 
 is difficult to be 
 the warm dark matter 
 which can form the large cosmological 
 structure\cite{gravitinobound}. 
Another 
 dark matter, such as axion, might 
 be needed for the large structure. 
}  
Anyhow,  
 the FCNCs from 
 the possible Planck suppressed operators in 
 the gravity mediation\cite{HPN} 
 are negligible.

As for the mass of $\Phi$, 
 the scalar component 
 is estimated as 
 $m^2/\Lambda\sim 3.2$ GeV from Eqs.(\ref{26}) 
 while 
 the fermion component is 
 $[F_\Phi^\dag \Phi^\dag/\Lambda^2]
 _{\bar{F}}\sim 10$ keV from Eq.(\ref{kahler2}). 
However, we should notice that 
 there are one- (two-) loop diagrams\footnote{
They are similar to the ordinary gauge mediation 
 diagrams, where $\rho$ and $\tilde{\rho}$ 
 propagate in the loops\cite{CST}. 
} 
 which 
 lift up fermion (scalar) masses of both 
 $\Phi$ and $\Phi_0$ as 
 $10^4$ GeV. 
Thus, the 
 mass spectra of the ISS fields are summarized in 
 the following table. 
\begin{eqnarray}
\begin{array}{|c||c|c|}
\hline
\hbox{Fields}   & \hbox{fermion mass} & \hbox{scalar mass} \\  
\hline
\hline
\rho,\tilde{\rho} & \Phi \sim 10^7 \hbox{ GeV}  &  
\Phi \sim 10^7 \hbox{ GeV} \\
\hline
\hat{Y},Z,\tilde{Z},\chi,\tilde{\chi} &
 m\sim  10^{6.5} \hbox{ GeV}  & m\sim  10^{6.5} \hbox{ GeV}  \\
\hline
\hat{\Phi} & 
 0.01\times \frac{F_\Phi}{\Phi} \sim  10^{4} \hbox{ GeV}  &
 0.01\times \frac{F_\Phi}{\Phi} \sim  10^{4} \hbox{ GeV}  \\
\hline
\hline
\hbox{gaugino} &
  10^{2} \hbox{ GeV}  & -  \\
\hline
\hbox{sfermion} &
- & 10^{4} \hbox{ GeV}   \\
\hline
\end{array}
\nonumber 
\end{eqnarray}
Here we take $m \sim \tilde{m}$, for simplicity.        


We should notice that 
 the strong-coupling scale is about
 $\Lambda\simeq 10^{12.5}$ GeV in Eq.(\ref{29}), 
 which is far below the GUT scale, 
 $2\times 10^{16}$ GeV. 
Thus the model, unfortunately, cannot 
 trace the gauge coupling running. 
For future reference we will show the QCD renormalization group 
 equation (RGE) of    this model 
 in Appendix B.

\section{Summary and discussions}

We have analyzed the ISS SUSY breaking 
 model 
 in the SUGRA framework. 
This is a simple and natural setup which demands 
 neither extra superpotential interactions nor 
 an additional gauge symmetry.  
In the SUGRA setup, 
 the $R$-symmetry is explicitly 
 broken by the constant term 
 in the superpotential, and 
 pseudo-{moduli} field  
 naturally takes non-zero VEV 
 through  a vanishing cosmological constant 
 condition. 
Sfermions tend to be heavier than 
 gauginos, and 
 the strong-coupling scale is 
 determined once the ratio of 
 sfermion 
 to gaugino masses 
 is fixed. 
The meta-stable SUSY 
 breaking vacuum can be sufficiently
 long-lived as shown in Appendix A. 

As for the $\mu$-term which also breaks 
 the $R$-symmetry, it could be derivable in the SUGRA 
 setup through the 
 Giudice-Masiero mechanism\cite{Giudice:1988yz}.
However, it is too small in the parameter
 set of Eq.(\ref{29}), so that we need 
 another mechanism to produce 
 the suitable magnitude of $\mu$-term.\footnote{ 
A simple example of generating $\mu$-term is to introduce       
 a gauge singlet field\cite{NMSSM}. 
}

Finally, we comment on 
 the small magnitude of 
 Eq.(\ref{g24}). 
Remind that 
 this determinant 
 takes 
 non-zero value in 
 the SUGRA setup with 
 the 
 next leading K\"{a}hler potential, while  
 it vanishes 
 in the original ISS model. 
However, unfortunately, 
  it was too small. 
Can we find 
 another meta-stable vacuum? 
One candidate is near 
 a singular point of
 the K\"{a}hler potential, 
 $Y\sim \Lambda$.
This minimum  has nothing to do with 
 the SUGRA effects, and 
 the SUSY mass of $Z\tilde{Z}$ 
 is given by 
 $[\Phi^\dag Y^\dag/\Lambda^2]_{\bar{F}} \simeq \tilde{m}^2 / \Lambda$. 
In this case 
 the gaugino masses become 
$$
M_{\lambda_i} \sim N
 {\alpha_i \over 4 \pi}{\tilde{m}^2\over \Lambda}, 
$$
which 
 are 
 the same order as the
 sfermion masses. 
But,  
 this estimation  
 might not be reliable, 
 since 
 the VEV of 
 $Y$ should be smaller than $\Lambda$ 
 for the correct magnetic 
 description of this theory.


\vspace{5mm}
\leftline{\bf Acknowledgments}

N.H. would like to thank N. Maru and K. Oda 
 for a partial collaboration of the potential
 analysis in an early stage of this work. 
N.H. would also like to
 thank K. Yoshioka, 
 T. Yamashita, N. Okada,  S. Matsumoto, 
 Y. Hyakutake, S. Shimasaki, and   T. Ishii
 for useful discussions.  
N.H. is supported by the Grant-in-Aid for Scientific Research, Ministry
 of Education, Science and Culture, Japan (No.16540258 and No.17740146).

\appendix
\section{Stability of meta-stable vacuum}

Here let us check whether the SUSY breaking
 meta-stable vacuum (which is found in Section 4)
 is long-lived 
 than the age of universe or not.  
The true vacuum exists 
 at which $q,\tilde{q}$ are decoupled and
 the gaugino condensation occurs through the
 pure SUSY $SU(N_f-N_c)$ gauge theory. 
Neglecting $N_c, N_f$ factors and ${\cal O}(1)$ 
 coefficients, 
 the $F$-flat conditions and 
 a matching condition
 between $\Lambda$ and gaugino condensation 
 scale
 derive 
\bea
 \Phi  \simeq 
{m^{2(N_f-N_c)\over N_c}}  
 \Lambda^{3N_c-2N_f \over N_c},
\;\;\;\;\;\; 
 Y  \simeq 
{\tilde{m^2}\over m^{2(2N_c-N_f)\over N_c}} \Lambda^{3N_c-2N_f \over N_c},
\;\;\;\;\;\; q=\tilde{q}=0. 
\label{SUSYvac1}
\eea
Here
 we neglect small corrections of  
 ${\cal O}(M_P^{-1})$, 
 and then the potential height of this vacuum is 
 estimated as  
\bea
V_{\rm (SUSY)} \simeq -\tilde{m}^4 .
\label{SUSYvac2}
\eea

The distance between the false vacuum and
 the true vacuum is roughly estimated as 
\bea
\Delta \Phi \simeq 
 \left\{
\begin{array}{c}
m^{2(N_f-N_c)\over N_c} \Lambda^{3N_c-2N_f \over N_c} 
\;\;\;\;\;\;(N_c < N_f < 1.46 N_c), \\
\frac{\Lambda^2}{M_P}~\;\;\;\;\;\;\;\;\;\;\;\;\;\;\;\;\;\;\;\;\;\;\;\;
(1.46N_c <N_f < {3\over2}N_c), \\     
\end{array}
\right. 
\eea
where we take $m\sim\tilde{m}$ 
  ($\Phi\sim Y$) and Eq.(\ref{29}). 
The
 potential height of the local maximal\footnote{
A 
 local maximum is located at 
$\chi = \tilde{\chi} = 0$, 
$\Phi  
\simeq -{\sqrt{N_c}\Lambda^2\over 2\sqrt{3}\lambda M_P}$, 
and 
$Y 
\simeq
 -{\sqrt{N_c}\tilde{m}^2\Lambda^2\over 2\sqrt{3}(\lambda+2\lambda')m^2M_P}$.  
} 
 is of order 
\bea
V_{{\rm peak}} \simeq \tilde{m}^4. 
\eea
Then,  
 the bounce action in the triangle approximation
 is estimated as
\bea
\label{S}
S \sim \frac{(\Delta \Phi)^4}{V_{{\rm peak}}} 
\sim 
\left\{
\begin{array}{c}
\left({m\over\tilde{m}}\right)^4
\left(
\frac{\Lambda}{m} \right)^{4(3N_c - 2N_f)\over N_c} 
 \;\;\;\;\;\;(N_c < N_f < 1.46 N_c), \\
\left(\frac{\Lambda^2}{M_P} \right)^4{1\over \tilde{m}^4} 
\;\;\;\;\;\;\;\;\;\;\;\;\;\;\;\;
(1.46N_c <N_f < {3\over2}N_c). \\
\end{array}
\right. 
\eea

A lifetime of the meta-stable vacuum 
 is estimated as\cite{CST} 
\beq
\tau \sim {1\over \tilde{m}}
\left({1s\over 10^{24} \hbox{ GeV}^{-1}}\right)
\sqrt{2\pi\over S}\; e^{S},
\eeq
which mean that  
 $S\geq 113$ is required
  for 
 $\tau$
  to be longer than 
 the age of the universe, 
 $\tau_0 \sim 4.7\times 10^{17} \; s$. 
We can easily show that 
 $\tau$ is much longer than $\tau_0$  
 when $N_f<[{3\over2}N_c]$
  (the largest integer smaller
 than ${3\over2}N_c$) in 
 $N_c < N_f < 1.46 N_c$.    
As for 
 the cases of 
 $1.46N_c <N_f < {3\over2}N_c$ or 
 $N_f=[{3\over2}N_c]$
 in $N_c < N_f < 1.46 N_c$ \footnote{
The case of   
 $N_c=11$ and $N_f=16$ 
 induces  
 the most   stringent bound as $S\sim 152$
 in 
 $N_c < N_f < 1.46 N_c$.  
}, 
 $S={\cal O}(100)$ so that 
 more accurate estimations must be required 
 by taking $N_c, N_f$ factors and ${\cal O}(1)$ 
 coefficients 
 into account. 
Anyhow  the meta-stable vacuum 
  can be 
 sufficiently long-lived in 
 the large range of the parameter space.  


\section{RGE of the model}

Let us show 
 the one-loop RGE 
 of the QCD gauge coupling, that is given by 
\bea
\alpha_i^{-1}(\mu) = \alpha_i^{-1}(\mu') + \frac{b_i}{2\pi}
\ln \left(\frac{\mu}{\mu'} \right),\nonumber
\eea
where $b_i$ is one-loop beta function coefficient of the gauge group.  
The energy dependence of $b_3$ 
 is listed below. 
\begin{eqnarray}
\begin{array}{|c||c|c|}
\hline
\hbox{energy} 
  & b_3 \;(SU(N_c) \supset  \hbox{SM})
  & b_3 \;(SU(N_f-N_c) \supset  \hbox{SM}) \\
\hline
\hline
M_Z < \mu < 10^2 \hbox{ GeV}
 & b_3^{{\rm SM}} = 7  & b_3^{{\rm SM}} = 7 \\
10^2 \hbox{ GeV} < \mu < 10^4 \hbox{ GeV}
 &  b_3^{{\rm SM}} - \frac{2}{3} \times 3 = 5 
 &  b_3^{{\rm SM}} - \frac{2}{3} \times 3 = 5 \\
10^4 \hbox{ GeV }<\mu < 10^{6.5} \hbox{ GeV }
 & b_3^{{\rm MSSM}}-b_3^{\hat{\Phi}} = 0 
 & b_3^{{\rm MSSM}} = 3 \\ 
10^{6.5} \hbox{ GeV }<\mu < 10^{7} \hbox{ GeV }
 &  - (N_f - N_c)  
 &  - N_c - b_3^{\hat{Y},\chi,\tilde{\chi}}\times3 =-N_c-9 \\ 
10^{7} \hbox{ GeV }<\mu < \Lambda
 &  -2(N_f - N_c) 
 &  -2 N_c - 9  \\
\hline
\end{array}
\nonumber 
\end{eqnarray}
Here $b^{{\rm SM}}$ and $b^{{\rm MSSM}}$ are
 the QCD one-loop beta function 
 coefficients for the SM and the MSSM, 
 respectively. 
%
Taking 
 $\alpha_3(M_Z)^{-1} \sim 8.47$        
and $b_3$ listed  above, 
 the QCD coupling
 at $\Lambda$ 
 is estimated as 
\bea
\hspace*{-5mm}&& 
\alpha_3(\Lambda)^{-1} 
\sim 
8.47 + \frac{1}{2\pi}
\left[
7\ln\left({M_\lambda\over M_Z}\right)
+5\ln\left({m_{\tilde{f}}\over M_\lambda}\right)
-2(N_f-N_c)\ln\left({\Lambda\over m}\right)
\right],\nonumber
\eea
when the SM gauge group is embedded into 
 $SU(N_c)$. 
Here 
 we have 
 neglected  
 the mass difference between $\rho, \tilde{\rho}$ and
 $Y,Z,\tilde{Z},\chi,\tilde{\chi}$, and 
 then taken a more  stringent bound.       
If we require that
 the QCD coupling constant
 is perturbative at $\Lambda$ 
 ($\alpha_3(\Lambda) < 1$), 
 the constraint for the flavor number of messengers $(N_f-N_c)$ 
 should be $(N_f - N_c) < 2.4$. 
As for the SM gauge group is 
 embedded into $SU(N_f-N_c)$, the 
 $b_3$ 
 is estimated as 
\bea
&& 
\alpha_3(\Lambda)^{-1} 
\sim 
8.47 + \frac{1}{2\pi}
\left[
7\ln\left({M_\lambda\over M_Z}\right)
+5\ln\left({m_{\tilde{f}}\over M_\lambda}\right)
+3\ln\left({m \over m_{\tilde{f}}}\right) 
-(2N_c+9)\ln\left({\Lambda\over m}\right)
\right].\nonumber
\eea
This shows the QCD coupling blows up soon above 
 $m$ (around $10^{8}$ GeV for $N_c=11$\footnote{
The embedding of the SM gauge group into 
 $SU(N_f-N_c)$ requires
 $N_c > 10$\cite{KOO}.  
}).


\end{document}